\documentclass[12pt]{article}

\usepackage{multirow}
\usepackage{amsfonts}
\usepackage{amssymb}
\usepackage{amscd}
\usepackage{cite}
\usepackage{amsmath}
\usepackage{bbm}
\usepackage{hyperref}

\def\hybrid{\topmargin -20pt    \oddsidemargin 0pt
        \headheight 0pt \headsep 0pt
        \textwidth 6.25in       
        \textheight 9.5in       
        \marginparwidth .875in
        \parskip 5pt plus 1pt   \jot = 1.5ex}

\hybrid

\numberwithin{equation}{section}
\numberwithin{table}{section}

\newcommand{\comment}[1]{}

\newcommand{\be}{\begin{equation}}
\newcommand{\ee}{\end{equation}}
\newcommand{\M}{{\bf M}}
\newcommand{\bea}{\begin{eqnarray}}
\newcommand{\eea}{\end{eqnarray}}
\newcommand{\ba}{\begin{aligned}}
\newcommand{\ea}{\end{aligned}}

\def\cN{{\cal N}}
\def\cF{{\cal F}}
\def\cG{{\cal G}}

\def\nv{n_{\rm v}}
\def\nh{n_{\rm h}}

\renewcommand{\Im}{\operatorname{Im}}

\newcommand\iu{\operatorname{i}}
\newcommand\diff{\mathrm{d}}
\newcommand\rmi{\mathrm{i}}

\newcommand{\F}{F}

\def\nvp{\hat{n}_{\rm v}}
\def\ax{{\tilde \phi}} 

\def\Y{Y}

\def\Mpl{M_{\rm Pl}}

\begin{document}
\begin{titlepage}
\begin{center}
\rightline{\small ZMP-HH/13-6}
\vskip 1cm

{\Large \bf
The rigid limit of $N=2$ supergravity}
\vskip 1.2cm

{\bf  Bobby E. Gunara$^{a}$, Jan Louis$^{b,c}$, Paul
  Smyth$^{d}$,\\[1ex] 
Luca Tripodi$^{b,c}$ and Roberto Valandro$^{e,f}$}

\vskip 0.6cm

$^{a}${\em Faculty of Mathematics and Natural Sciences, Institut
Teknologi Bandung, Jl. Ganesha 10 Bandung, 40132, Indonesia}

\vskip 0.2cm

$^{b}${\em II. Institut f\"ur Theoretische Physik der Universit\"at Hamburg, Luruper Chaussee 149, 22761 Hamburg, Germany}
\vskip 0.2cm

{}$^{c}${\em Zentrum f\"ur Mathematische Physik,
Universit\"at Hamburg,\\
Bundesstrasse 55, D-20146 Hamburg}
\vskip 0.2cm

$^d$ {\em Institut de Th\'eorie des Ph\'enom\`enes Physiques,
EPFL,\\ CH-1015 Lausanne, Switzerland} 
\vskip 0.2cm

$^{e}${\em The Abdus Salam International Center for Theoretical Physics ICTP, \\ Strada Costiera 11, Trieste 34014, Italy}
\vskip 0.2cm

$^{f}${\em INFN, Sezione di Trieste, Italy}
\vskip 0.6cm

{\tt bobby@fi.itb.ac.id, jan.louis@desy.de, paul.smyth@epfl.ch, luca.tripodi@desy.de, rvalandr@ictp.it}

\end{center}

\vskip 0.8cm

\begin{center} {\bf ABSTRACT } \end{center}

\noindent
In this paper we review the rigid limit of $N=2$ supergravity coupled
to vector and hypermultiplets. In particular we show 
how the respective scalar field spaces reduce to their global
counterparts. In the hypermultiplet sector we focus 
on the relation between the local and rigid c-map.

\bigskip

\vfill

May 2013


\end{titlepage}



\section{Introduction}

Supergravity theories are mainly discussed as effective low-energy theories
of some ultraviolet complete fundamental theory such as string theory.
In this low-energy limit the heavy string modes are integrated out
and only massless and light modes are retained. However, gravitational
interactions of the light states are kept and thus the Planck scale~$\Mpl$ 
appears in
the effective low-energy Lagrangian. For some applications it is of
interest to  decouple gravity in a second step 
by taking the rigid limit $\Mpl\to\infty$. 
In many cases this is straightforward
but it can also be a confusing issue. 

In this paper we focus on $N=2$ supergravity coupled
to vector and hypermultiplets and study its rigid limit in a Minkowski background.
Both multiplets contain scalar fields and 
$N=2$ supersymmetry dictates how the geometry of their respective 
field spaces as well as the other
couplings in the low-energy effective theory have to behave 
in the rigid limit.
For vector multiplets the scalar geometry reduces from a 
projective special K\"ahler manifold (which is sometimes called a 
`local special K\"ahler manifold') to a
special K\"ahler manifold (which is sometimes called a 
`rigid special K\"ahler manifold').
For hypermultiplets the scalar geometry reduces from a 
quaternionic K\"ahler to a hyper-K\"ahler manifold.
This limiting procedure has previously been discussed in
Refs.~\cite{Bagger:1983tt,Galicki:1985qv,Seiberg:1994rs,Billo:1998yr,David:2003dh,Ambrosetti:2010tu}
(for reviews see, for example, \cite{Andrianopoli:1996cm,Freedman:2012zz}).

This paper is inspired by the situation where $N=2$ supergravity appears as
the low-energy limit of 
string theory, but our considerations hold for any UV theory 
with similar properties.
In string theory one typically has two different classes of light
scalar fields 
with different types of couplings in the effective theory. 
On the one hand, there are scalars (often denoted as moduli) which only couple gravitationally and whose background values 
can be as large as $\Mpl$. 
These fields are essentially frozen to their background values
with only harmonic fluctuations left.
On the other hand, one can have charged scalars
which can have gauge interactions as well as gravitational interactions.
In an unbroken 
gauge theory their background values are zero
and they contribute to non-trivial dynamics at low energies.
These scalars also can have a
non-zero background value which induces spontaneous symmetry
breaking at a scale set by the background values.
Here we do not specify the details of the gauge dynamics but we do 
distinguish scalar fields $\Phi$ with background values  
$\Phi_0={\cal  O}(\Mpl)$
and scalar fields $\varphi$ with background values  
$\varphi_0 \ll {\cal  O}(\Mpl)$. Furthermore,
if the effective theory has no additional scale $\Lambda<\Mpl$
(such as the QCD- or Seiberg-Witten
scale \cite{Seiberg:1994rs})
any scalar field space reduces to flat space. Therefore we 
allow for a generic  $\Lambda$ 
and perform the rigid limit in the presence of non-zero $\Lambda$.
In addition, we also assume throughout the paper that both supersymmetries are 
unbroken.

The paper is organized as follows. In Section~\ref{section:N=2}
we recall that without any additional scale $\Lambda<\Mpl$ in the
theory the rigid limit of the scalar field space is flat. 
In Section~\ref{section:vecmul} we take the rigid limit in the
vector multiplet sector and show how the rigid prepotential
characterizing the rigid special K\"ahler geometry is related to
the prepotential of the local special K\"ahler geometry.
In Section~\ref{section:hypmul} we consider the rigid limit in the
hypermultiplet sector. Here we focus on special quaternionic K\"ahler
manifolds
which arise at the tree-level of type~II compactifications and which
are
characterized by the (local) c-map.
We show that in the rigid limit these spaces reduce to
hyper-K\"ahler manifolds characterized by the rigid c-map \cite{Cecotti:1988qn,Ferrara:1989ik}.\footnote{These manifolds have also been of interest recently in
relation to wall-crossing phenomena. (For recent reviews see, for
  example, \cite{Moore:2012yp,Alexandrov:2013yva} and references therein.)}

%
\section{Preliminaries}
\label{section:N=2}

We shall first briefly recall the spectrum and couplings of four-dimensional ${\cal
N}=2$ supergravity (for a review see e.g.\ \cite{Andrianopoli:1996cm,Freedman:2012zz}). The theory consists of a gravitational multiplet, $\nv$ vector multiplets and $\nh$ hypermultiplets. The gravitational multiplet $(g_{\mu\nu},\Psi_{\mu {\cal A}}, A_\mu^0)$ contains the spacetime metric $ g_{\mu\nu}, \mu,\nu =0,\ldots,3$, two gravitini $\Psi_{\mu {\cal A}}, {\cal A}=1,2$, and the graviphoton $A_\mu^0$. A vector multiplet $(A_\mu,\lambda^{\cal A}, t)$ contains a vector $A_\mu$, two gaugini $\lambda^{\cal A}$  and a complex scalar $t$. Finally, a hypermultiplet $(\zeta_{\alpha}, q^u)$ contains two hyperini $\zeta_{\alpha}$ and 4 real scalars $q^u$. 
The bosonic Lagrangian is given by
\begin{equation}\begin{aligned}\label{sigmaint}
{\cal L}\ =\  &\tfrac{1}{2\kappa^2} R 
+\tfrac14 {\rm Im}\mathcal{N}_{IJ}(t,\bar t)\,F^{I }_{\mu\nu}F^{\mu\nu\, J}
- \tfrac18 {\rm Re}\mathcal{N}_{IJ}(t,\bar t)\,
\varepsilon^{\mu\nu\rho\sigma} F^{I}_{\mu\nu} F_{\rho \sigma}^J\\
&- g_{i\bar \jmath}(t,\bar t)\, D_\mu t^i D^\mu\bar t^{\bar \jmath}
-h_{uv}(q)\, D_\mu q^u D^\mu q^v - 
V(t,\bar t,q)
\ ,
\end{aligned}\end{equation}
where  $\kappa^{-1} =  8\pi\Mpl$.
We have chosen canonical mass dimension one
for the vector and scalar fields and thus 
the sigma-model metrics
$g_{i\bar \jmath}(t,\bar t),\, i,\bar\jmath = 1,\ldots,\nv$ and
$h_{uv}(q), u,v=1,\ldots,4\nh,$ together with
the kinetic matrix $\mathcal{N}_{IJ}(t,\bar t),\, I,J = 0,\ldots,\nv$   
are dimensionless couplings which we specify
in more detail in the following sections.
$V$ is the dimension four scalar potential.

Before we study the rigid limit of this theory
let us make some general observations.
The (dimensionless) space-time metric is expanded around a Minkowski background
$\eta_{\mu\nu}$ as
\be
g_{\mu\nu} = \eta_{\mu\nu} + \kappa h_{\mu\nu} +\ldots\ ,
\ee
while the scalar fields are expanded around their background values
$t_0$ and $q_0$
\be
t^i= t_0^i + \delta t^i\ ,\qquad q^u = q^u_0 + \delta q^u\ .
\ee
The couplings in \eqref{sigmaint} can be expanded for small fluctuations
accordingly
\be\ba\label{cexp}
\mathcal{N}_{IJ}(t,\bar t) &= \mathcal{N}_{IJ}(t_0,\bar t_0) +
 \partial_i\mathcal{N}_{IJ}(t_0,\bar t_0) \delta t^i +
\partial_{\bar i}\mathcal{N}_{IJ}(t_0,\bar t_0) \delta t^{\bar i} + \ldots\ ,
\\
g_{i\bar \jmath}(t,\bar t) &= g_{i\bar \jmath}(t_0,\bar t_0) +
\partial_kg_{i\bar \jmath}(t_0,\bar t_0) \delta t^k+
 \partial_{\bar k}g_{i\bar \jmath}(t_0,\bar t_0) \delta t^{\bar k}+ \ldots\ ,
\\
h_{uv}(q) &= h_{uv}(q_0) +\partial_w h_{uv}(q_0)\delta q^w+
\ldots\ . \ea\ee Since we have chosen canonical mass dimension one
for the scalar fields and the gauge bosons, the couplings
$\mathcal{N}_{IJ},g_{i\bar \jmath},h_{uv}$ are dimensionless while
their respective derivatives have mass dimension $-1$. Therefore, if
the theory under consideration has no scale other than 
$\Mpl$ then all higher order terms  in the expansions \eqref{cexp},
i.e. all terms including derivatives of the couplings, scale with
$\kappa$ and thus vanish in the rigid limit $\kappa\to0$. This implies
that for each quantity only the first term survives, where the couplings are
evaluated at the constant background values of the scalar fields.
At that point in field space all three matrices
$\mathcal{N}_{IJ},g_{i\bar \jmath},h_{uv}$ are constant and hence
can be diagonalized. Thus, in this situation the scalar field
space of the rigid theory is flat and the gauge kinetic matrix can be chosen
diagonal. Of course what we are recalling here is that 
without an intermediate scale any kinetic term reduces to its renormalizable form in the rigid
limit.

The situation changes if the theory has a second, possibly dynamical
scale  $\Lambda$
(such as the QCD scale or the Seiberg-Witten scale) which is well below
the Planck scale $\Lambda\ll\Mpl$.
In this case the above reasoning does not hold as the derivatives
of the couplings $\partial_kg_{i\bar \jmath}(t_0,\bar t_0)$ etc.\
do not necessarily scale with $\kappa$ but could instead have a $\Lambda^{-1}$ dependence
and thus do not have to vanish in the $\kappa\to0$ limit.
As a consequence there can be a non-trivial field space and
non-trivial gauge couplings in the rigid theory -- Seiberg-Witten
theory \cite{Seiberg:1994rs} and generalizations thereof
being a prominent example. It is this generic situation which we are concerned with in this paper.

In order to proceed we need to make one further generic distinction.
The theory under consideration can have two classes of scalar fields, which
for now we denote by $\Phi$ and $\varphi$. The $\Phi$'s  have
Planck-sized background values, i.e., $\Phi_0 = {\cal O}(\Mpl)$, and 
are typically flat directions (i.e., moduli) of the potential $V$.\footnote{In
  principle they could also
be fixed by  $V$, but in that case they would generically have
masses of order ${\cal O}(\Mpl)$ and thus would have been integrated
out of the low-energy theory.}
The fields $\varphi$ have  vanishing or small
background values $\varphi_0\ll\Mpl$.\footnote{We also allow for the case $\varphi_0>\Lambda$ but then insist on $\Lambda<\varphi_0\ll\Mpl$.}
Since all couplings in \eqref{sigmaint} are dimensionless they depend
on the ratios $\frac{\Phi}{\Mpl},\frac{\Phi}{\Lambda},\frac{\varphi}{\Mpl},
\frac{\varphi}{\Lambda}$. The dependence $\frac{\Phi}{\Lambda}$ cannot occur
as in this case the couplings would diverge,
while any term that depends on the ratio $\frac{\varphi}{\Mpl}$
disappears in the $\Mpl\to\infty$ limit.
Thus, in the rigid limit only the dependence on
$\frac{\Phi}{\Mpl}$ and $\frac{\varphi}{\Lambda}$ needs to be kept.
Furthermore, in that limit 
any derivative with respect to $\Phi$ necessarily scales
like $\Mpl^{-1}$ while
derivatives with respect to $\varphi$ scale
like $\Lambda^{-1}$. Hence, in the limit $\Mpl\to\infty$ the fluctuations
of $\Phi$ are suppressed, or frozen, and only fluctuations in $\varphi$ survive.
Schematically we thus have
\be
g(\Phi,\varphi) = g(\Phi_0,\varphi)
= g(\Phi_0,\varphi_0) + \partial_\varphi g(\Phi_0,\varphi_0)
\delta\varphi +\ldots\ ,
\ee
i.e., we can replace $\Phi$ by its background value $\Phi_0$ while
we keep $\varphi$ as a dynamical field.
Thus $\Phi$ can be viewed as part of a hidden sector while $\varphi$
denotes the observable sector.

Let us stress that the distinction between $\Phi$ and
$\varphi$ usually can only be maintained locally.
It is often the case that
at specific subspaces in the field space heavy modes, which
have been integrated out, become light. This effect generically shows up
as a singularity in some couplings.
In this case the choice of low-energy excitations
is no longer consistent and one has to go to a different effective theory
with different field variables and couplings.
In the following we will always assume that we work in a region of the
moduli space where the distinction between $\Phi$ and $\varphi$ is
unambiguous.
This is also consistent with the fact
that taking the rigid limit zooms-in on a specific
region of the moduli space. Only the points in the neighbourhood of
such a region are kept in the rigid scalar field space, i.e. a generic
point is at distance $M_{\rm pl}$ from this region and it will be sent
to infinite distance in the rigid limit.

\section{Vector multiplet sector}
\label{section:vecmul}

Let us start with the vector multiplet sector. In this case the metric $g_{i\bar \jmath}$ in \eqref{sigmaint}
is defined on the $2\nv$-dimensional special K\"ahler manifold ${\M}_{\rm v}$ \cite{deWit:1984pk,Craps:1997gp}.
This means that $g_{i\bar \jmath}$ obeys
\begin{equation}\label{gdef}
g_{i\bar \jmath} = \partial_i \partial_{\bar \jmath} K^{\rm v}\ ,
\qquad \textrm{for}\qquad
K^{\rm v}= -\kappa^{-2}\ln\iu\Y\ ,\qquad
\Y= \kappa^2\left( \bar X^I \F_I - X^I\bar \F_I \right)\ ,
\end{equation}
where both $X^I(t)$ and $\F_I(t)$, $I= 0,1,\ldots,\nv$, are dimension-one
holomorphic functions of the scalars $t^i$.  
$\F_I = \partial\F/\partial{X^I}$ is the derivative of a holomorphic
prepotential $\F(X)$  which is homogeneous of degree two in $X$. 
Furthermore,
it is possible to go to a system of `special coordinates' where 
$X^I= (\Mpl,t^i)$ or more generally $t^i = \Mpl\,\frac{X^i}{X^0}$.
Due to the homogeneity property of $F(X)$ it is convenient to define
a dimensionless function $\cF$ by 
$F=(X^0)^2\cF(X^i/X^0) =\kappa^{-2}\cF(t^i)$.
In terms of $\cF$ the K\"ahler potential \eqref{gdef} reads
\be\label{KcF}
K^{\rm v}= -\kappa^{-2}\ln\iu\Y\ ,\qquad
\Y= 
2(\cF-\bar\cF) - (t-\bar t)^i (\cF + \bar \cF)_i \ .
\ee
The kinetic matrix $\mathcal{N}_{IJ}$ in \eqref{sigmaint} is also a function of the $t^i $ given by
\begin{equation}
  \label{Ndef}
  {\cal N}_{IJ} = \bar \F_{IJ} +2\iu\ \frac{\mbox{Im} \F_{IK} X^K \mbox{Im}
    \F_{JL} X^L}{ X^L\mbox{Im} \F_{LK}  X^K} \ ,
\end{equation}
where $\F_{IJ}=\partial_I \F_J$. 
(See e.g., \cite{Craps:1997gp} for further details).

In terms of the set-up of the previous section we now
need to distinguish between $\Phi$ and $\varphi$ for the vector
multiplets. Let us denote 
all scalar fields in vector multiplets with
Planck-sized background values by $\Phi$, where we suppress the index for notational
simplicity. By a slight abuse of notation we will
continue to label the subset of $\nvp$ scalars in the observable sector by
$t^i, i=1,\ldots,\nvp$. 

Before any expansion we can choose the prepotential to be of 
the generic form 
\be \label{Fexp}
\cF= \cF^\Phi(\kappa\Phi) + \cF^t(\kappa\Phi,t^i,\kappa,\Lambda)\ ,
\ee 
where  $\cF^t(\kappa\Phi,t^i=0,\kappa,\Lambda)=0$ holds.\footnote{Prepotentials of this form do not allow for terms like $e^{t/M_{\rm Pl}}$, for example. However, when such a term is expanded around $\kappa\sim 0$, it has actually a term constant in $t$ and one term vanishing at $t=0$. Hence, since at the end we are interested in the limit $\kappa\rightarrow 0$, we can start for our generic considerations from prepotentials of the form \eqref{Fexp}.} With this choice
the first term $\cF^\Phi(\kappa\Phi)$ encodes the geometry of the hidden sector,
and since $\cF^\Phi$ is
dimensionless it only depends on the product $\kappa\Phi$.
The second term
$\cF^t$
captures any non-trivial gauge dynamics of the observable sector
(and thus depends on $\Lambda$ and $t$) and the 
interactions between hidden and observable sector, or in other words 
between $\Phi$ and $t$. The condition $\cF^t(\kappa\Phi,t^i=0)=0$
says that any field independent term is chosen to be part of $\cF^\Phi$.

Let us first focus on $\cF^\Phi$ and expand 
$\Phi = \Phi_0+\delta\Phi$. For small $\delta\Phi$ this implies the
Taylor expansion
\be\label{Fphikappa}
\cF^\Phi = \cF^\Phi( \kappa\Phi_0) + \kappa\partial\cF^\Phi( \kappa\Phi_0)\,\delta\Phi
+ \tfrac12\kappa^2\partial^2\cF^\Phi( \kappa\Phi_0)\,\delta\Phi^2+{\cal O}(\kappa^3)
\ , \ee
with the first term being constant.
Note that we are
assuming $\Phi_0={\cal O}(\Mpl)$ and thus $\kappa\Phi_0$ is a
dimensionless parameter.
Inserting this into the definition of $K^{\rm v}$ \eqref{KcF},
 expanding the $\ln$ and ignoring $\cF^t$ for the
moment yields  \cite{Billo:1998yr}
\begin{equation}\label{eq:Kr-g}
K^{\rm v}= -\ln Y(\kappa\Phi_0) + Y^{-1}(\kappa\Phi_0)\, K^{\rm v}_{\rm r}
+{\cal O}(\kappa)\ ,
\ee
where
\be \label{Kgrav}
K^{\rm v}_{\rm r} = h(\delta\Phi)+\bar h(\delta\bar\Phi) 
+ g_{\Phi\bar\Phi}(\kappa\Phi_0)\,\delta\Phi\delta\bar\Phi 
\ .
\ee 
$h(\Phi)$ is holomorphic in $\delta\Phi$
and thus does not enter the K\"ahler metric. 
Note that the $\ln Y(\kappa\Phi_0)$ term and the
normalization factor $Y^{-1}(\kappa\Phi_0)$ in \eqref{eq:Kr-g}
are constant and thus can be absorbed by redefining $h$ and $g_{\Phi\bar\Phi}$,
or $\delta\Phi$ in \eqref{Kgrav}. We have chosen to display the rigid limit in the form 
\eqref{eq:Kr-g} for later convenience and 
to stress that a non-zero $Y(\kappa\Phi_0)$ 
is essential in order to expand the $\ln$.
As anticipated, only the quadratic term of $K^{\rm v}_{\rm r}$
contributes in the $\kappa\to 0$ limit,  leading to a
constant (and thus flat) metric $g_{\Phi\bar\Phi}=g_{\Phi\bar\Phi}(\kappa\Phi_0)$ 
which can be computed straightforwardly 
in terms of $\cF^\Phi$. The precise expression is not very illuminating and, furthermore, by
an appropriate redefinition of the $\delta\Phi$ one can
always choose $Y^{-1} g_{\Phi\bar\Phi}=\delta_{\Phi\bar\Phi}$.

Let us now turn to the second piece  $\cF^t(\kappa\Phi,t^i,\kappa,\Lambda)$ in
\eqref{Fexp}.
First of all this term can in principle induce (constant) corrections
to  the metric $g_{\Phi\bar\Phi}$ depending on $\frac{t_0^i}{\Lambda}$.
As such corrections can always be absorbed into a redefinition of
$\delta\Phi$ we will not consider them in the following. 
As we argued in the previous section we can
replace $\Phi$ by its background value and only consider
  $\cF^t(\kappa\Phi_0,t^i,\kappa,\Lambda)$.\footnote{We argued in the
    previous section that a dependence on the ratio $\frac{\Phi_0}{\Lambda}$
cannot occur as it would diverge in the $\Mpl\to\infty$ limit and thus
also $\cF^t$ can only depend on $\kappa\Phi$. 
Replacing $\Phi$ by its background value 
then means that we are neglecting all derivatives of $\cF^t$ with
respect to $\Phi$ as they disappear in the rigid limit.}
It is convenient to make the $\kappa$-dependence explicit by defining
\be\label{Fkappa}
\cF^t(\kappa\Phi_0,t^i,\kappa,\Lambda) =
\sum_{n=0}^\infty \kappa^n \cF^{t(n)}(\kappa\Phi_0,t^i,\Lambda)\ ,
\ee
where the $\cF^{t(n)}$  have mass dimension $n$.  For dimensional reasons
$\cF^{t(0)}$ has to be constant and with our
convention $\cF^t(\Phi,t^i=0,\kappa,\Lambda)=0$ we have chosen
this constant to be part of $\cF^\Phi(\Phi_0)$, implying that
$\cF^{t(0)}=0$. Similarly,
$\cF^{t(1)}$ would be linear in $t$ and so violate gauge
invariance. More importantly, this linear term would correspond to terms of the form $F\sim X^0 X^i$ and/or
$F\sim X^\Phi X^i$
in the homogenous  degree-two prepotential $F$ 
appearing in \eqref{gdef} and
\eqref{Ndef}. This
violates our assumptions that the $t^i$ decouple
from the hidden sector and, as we will see shortly, 
would lead to mixing with the graviphoton.
Thus, consistency constrains the expansion \eqref{Fkappa} and requires
\be
 \cF^{t(0)}= \cF^{t(1)} =0\ .
\ee
It is then straightforward to see that 
after expanding the $\ln$ in \eqref{KcF} and taking 
the limit  $\kappa\to 0$ only 
$ \cF^{t(2)}(\kappa\Phi_0,t^i,\Lambda)$ survives.
Inserting this into \eqref{KcF} and expanding the $\ln$, and including the
$\Phi$-dependence given in \eqref{Kgrav},
we obtain \eqref{eq:Kr-g} with
\be\label{eq:Kr-r}
K^{\rm v}_{\rm r}=h(\delta\Phi,t)+\bar h(\delta\bar\Phi,\bar t)
-\iu\big({\delta\bar{\Phi}} \cF^{\Phi(2)}_\Phi- \delta\Phi\bar\cF^{\Phi(2)}_{\bar\Phi}\big)
- \iu \big({\bar{t}}^i \cF^{t(2)}_i- t^i\bar\cF^{t(2)}_{\bar i}\big)\ .
\end{equation}
where we have defined
$ \cF^{\Phi(2)}=\tfrac{i}4g_{\Phi\bar\Phi}(\kappa\Phi_0)\,\delta\Phi^2$ and  
$h(\Phi,t)$ is again a holomorphic function which does not enter the metric.
Indeed \eqref{eq:Kr-r} is the K\"ahler potential of a rigid special
K\"ahler manifold with the metric \cite{Freedman:2012zz}
\be\label{eq:g}
g=\left(\begin{array}{cc}
g_{\Phi\bar\Phi}&0\\
0& g_{i\bar j}
\end{array}\right) =
2\left(\begin{array}{cc}
{\rm Im}\cF^{\Phi(2)}_{\Phi\Phi}&0\\
0& {\rm Im}\cF^{t(2)}_{i\bar j}
\end{array}\right)\ ,
\ee
so that $\cF^{\Phi(2)}+\cF^{t(2)}$ can be identified as the rigid prepotential.
Note that these standard relations only hold up to
the constant normalization factor $Y^{-1}(\kappa\Phi)$
which, however,  can be absorbed into the rigid prepotential or 
by a field redefinition of $\delta\Phi$ and $t$.
Let us also reiterate that a non-zero $Y^{-1}(\kappa\Phi)$ is essential
for a consistent expansion of the $\ln$.

Let us now turn to the gauge kinetic matrix defined in \eqref{Ndef}.
It parameterizes the gauge couplings of the graviphoton  together with
the gauge bosons in the vector multiplets. Since we distinguish the
purely gravitationally coupled scalars $\Phi$ and the observable
scalars $t^i$ we need to make the same distinction for the gauge bosons
as they reside in the same multiplet. Thus $A_\mu^0$ denotes the graviphoton,
$A_\mu^\Phi$ the gauge bosons which form vector multiplets with the scalars
$\Phi$ and $A_\mu^i$ the gauge bosons which form vector multiplets
with the scalars $t^i$.  In the rigid limit the graviphoton $A_\mu^0$ can mix with the gravitationally coupled $A_\mu^\Phi$ but there should be no couplings with the observable $A_\mu^i$. In other words, ${\cal N}_{IJ}$ should be block-diagonal with ${\cal N}_{0i}={\cal N}_{\Phi i}=0$ in the rigid limit. Let us now explicitly check the consistency of this requirement with the constraints on the prepotential that we discussed above. First, we observe
that both $F_{0i}\sim {\cal O}(\kappa) $ and $F_{\Phi i}\sim {\cal O}(\kappa)$,  as the derivative with respect $t^i$ (or rather $X^i$) means that
only the Planck suppressed
$\cF^{t(2)}$ contributes. For the components ${\cal N}_{0i}$ and ${\cal N}_{\Phi i}$ the non-holomorphic second term in \eqref{Ndef} is suppressed because in both case it contains the Planck suppressed $\cF^{t(2)}$ in the numerator. Hence we see that ${\cal N}_{0i}\sim{\cal O}(\kappa)$ and ${\cal N}_{\Phi i}\sim{\cal O}(\kappa)$, and so the gravitational and visible sectors decouple. The components ${\cal N}_{00}, {\cal N}_{0\Phi}, {\cal N}_{\Phi\Phi}$, on the other
hand, are constant at leading order as they depend on $\cF^\Phi$.
Finally, in ${\cal N}_{ij}$ the non-holomorphic second term is Planck suppressed,
as the numerator only depends on $\cF^{t(2)}$, while the denominator
can be $ {\cal O}(1)$.
Thus we have
\be\label{tau}
{\cal N}_{ij} = \bar F_{ij} =\bar \cF^{t(2)}_{ij}\ ,
\ee
implying that ${\rm Im}{\cal N}_{ij} =-{\rm Im} \cF^{t(2)}_{ij} =- \tfrac12g_{ij}$,
which is indeed the correct relation in global $\cN=2$ supersymmetry \cite{Freedman:2012zz}.

Let us close this section with two
 explicit examples. As a first, simple example we will consider the prepotential
\be
F=\tfrac{\iu}{4}\big((X^0)^2 - \eta_{ij}X^iX^j\big)\ ,\qquad i=1,\ldots,\nv\ ,
\ee
with $\eta_{ij}$ real. Inserted  into  \eqref{gdef} or  \eqref{KcF}
and using $ t^i = \kappa^{-1}\frac{X^i}{X^0}$  yields
\be
\cF = \tfrac{\iu}{4}(1-  \kappa^2 \eta_{ij}t^it^j)\ ,\qquad
K=-\ln\big(1-\kappa^2 \eta_{ij}t^i\bar t^j\big) + {\rm const.}\ ,
\ee
which is the K\"ahler potential of the space
$\frac{SU(1,\nv)}{U(1)\times SU(\nv)}$.\footnote{Note that we could add
  terms of the form $X^0X^i$ in $F$ which for a purely quadratic $F$ 
can always be rotated away. Furthermore, any imaginary part of $\eta_{ij}$
does not contribute to $K$ (and thus the metric)
 but does contribute to the $\theta$-angle as can be seen from \eqref{tau}.}
In terms of the notation \eqref{Fkappa} we infer
$\cF^{(2)}=-\tfrac{\iu}{4} \eta_{ij}t^it^j$ and inserting this into
\eqref{eq:g} we obtain the flat metric $g_{i\bar j} = \eta_{ij}$.

As a second example we consider the prepotential
\be
F=   \frac{\iu}{4}\frac{X^1}{X^0}(X^2X^3 - \eta_{ij} X^iX^j)\ ,\qquad
i,j=4,\ldots,\nv\ ,
\ee
with $\eta_{ij}$ again real.
Inserting this into  \eqref{gdef} or  \eqref{KcF}
and using 
\be
S= \kappa^{-1}\frac{X^1}{X^0}\ ,\quad T=
\kappa^{-1}\frac{X^2}{X^0}\ ,\quad
U=\kappa^{-1}\frac{X^3}{X^0}\ ,\quad 
t^i = \kappa^{-1}\frac{X^i}{X^0}
\ee  
yields 
\be\label{FSTUt}
\cF = \frac{\iu}{4}\kappa^3 S(TU- \eta_{ij}t^it^j)\ ,
\ee 
and 
\be
K=-\ln(S-\bar S)-\ln\big((T-\bar T)(U-\bar U)- \eta_{ij}(t-\bar
t)^i (t-\bar t)^j\big) + {\rm const.}\ ,
\ee
which is the K\"ahler potential of the space

\be
\frac{SU(1,1)}{U(1)}\times \frac{SO(2,\nv-1)}{SO(\nv-1)\times SO(2)}\ .
\ee
Here $S,T$ and $U$ are scalars of gravitationally coupled vector
multiplets, while the $t^i$ can be part of the observable sector.
In the spirit of this paper $S,T$ and $U$ are assumed to
have Planck-sized background
values $S_0,T_0,U_0$
and thus $\cF^{t(2)}=-\tfrac{\iu}{4} \kappa S_0 \eta_{ij}t^it^j$. Inserted into
\eqref{eq:g} we obtain the flat metric $g_{i\bar j} = \kappa S_0 \eta_{ij}$.
If the $t^i$ parameterize the Coulomb-branch of a non-Abelian gauge
theory then $\cF^{t(2)}$ is corrected at one-loop
and non-perturbatively. For $SU(2)$ with one modulus $t$
one finds \cite{Seiberg:1994rs}
\be\label{FSW}
\cF^{t(2)}=-\tfrac{\iu}{4}\kappa S_0 t^2 +
t^2\ln\frac{t^2}{\Lambda^2}
+ t^2\, \sum_{k=1}^\infty \cF_k\, \Big(\frac{\Lambda}{t}\Big)^{4k}\ .
\ee
The generalization to arbitrary gauge groups and the derivation of 
$\cF^{t(2)}$ from
string theory is reviewed in \cite{Lerche:1996xu}. 

Before turning to the hypermultiplet sector let us also note
that at special points in the $T-U$ plane a non-Abelian gauge
enhancement can occur.
For example, near $T\approx U$ one observes the enhancement
$U(1)^2\to SU(2)\times U(1)$. This is precisely the situation that we
mentioned above, in that on a subspace of the
moduli space additional states become light which change the effective 
description. In this case it is convenient to 
introduce the variables $T_\pm = \tfrac12 (T\pm U)$ 
such that near $T\approx U$ the background value of $T_-$ is small,
as is assumed for the $t^i$. In other words, near $T\approx U$
we should treat $T_-$ as an observable scalar field.
Inserted into \eqref{FSTUt} this yields
\be
\cF = \frac{\iu}{4} \kappa^3 S(T_+^2- T_-^2 - \eta_{ij}t^it^j)\ ,
\ee 
which displays the similarity of $T_-$ and the $t^i$.
The full perturbative and non-perturbative corrections of the $STU$-model
in string theory were derived in
\cite{deWit:1995zg,Antoniadis:1995ct,Kaplunovsky:1995tm,Klemm:1995tj,Harvey:1996gc},
while the rigid limit yielding the rigid prepotential \eqref{FSW} was 
explicitly performed in \cite{Kachru:1995fv}.



\section{Hypermultiplet sector}\label{section:hypmul}

\subsection{Generic case}

Let us now consider the rigid limit of the hypermultiplet geometry.
In this sector $h_{uv}(q)$ appearing in the Lagrangian
\eqref{sigmaint}
is constrained to be the metric of a  ($4\nh$-dimensional)
quaternionic K\"ahler manifold ${\M}_{\rm h}$
\cite{Bagger:1983tt,deWit:1984px,Freedman:2012zz}. 
Such manifolds have 
holonomy group $Sp(1)\times Sp(\nh)$ and 
they  admit a triplet of complex structures 
$J^x, x=1,2,3$, which satisfy the quaternionic algebra
$J^x J^y = -\delta^{xy}{\bf 1} + \epsilon^{xyz} J^z$. The metric
$h_{uv}$ is Hermitian with respect to all
three of complex structures. The associated  hyper-K\"ahler two-forms given by
$K^x_{uv} = h_{uw} (J^x)^w_v$ are covariantly closed with
respect to the $Sp(1)$ connection~$\omega^x$, i.e., 
$\nabla K^x \equiv dK^x + \epsilon^{xyz} \omega^y \wedge K^z=0$.
This implies that $K^x$ can be viewed as
$Sp(1)$ field strength of $\omega^x$ given by
\begin{equation} \label{def_Sp(1)_curvature}
 K^x =\diff \omega^x + \tfrac12 \epsilon^{xyz} \omega^y\wedge \omega^z\ .
\end{equation}

In the rigid limit, on the other hand,
global $N=2$ supersymmetry constrains $K^x$ to be  closed and 
the metric to be
hyper-K\"ahler with holonomy $Sp(n)$ \cite{AlvarezGaume:1981hm}.
Hyper-K\"ahler manifolds are Ricci-flat while quaternionic K\"ahler
manifolds are Einstein.
It was shown in \cite{Bagger:1983tt} that the $Sp(1)$ part of the
curvature scales with $\kappa^2$ and
the Riemann curvature tensor of a quaternionic K\"ahler manifold
decomposes as
\begin{equation}\label{qRiem}
R_{stuv} = \kappa^2 \hat R_{stuv} + W_{stuv}\ ,
\end{equation}
where $\hat R_{tsuv}$ is the (dimensionless) $Sp(1)$ curvature while 
$W_{stuv}$ is the Ricci-flat Weyl-curvature of a hyper-K\"ahler
manifold.\footnote{See \cite{Freedman:2012zz}
for a review.}
In terms of the metric $h_{uv}$ this again implies that $h_{\Phi\Phi}$ is flat for the purely
gravitationally coupled scalars, while one can have a non-trivial hyper-K\"ahler metric in an
observable sector if there is an additional scale $\Lambda$. It is difficult to make further, general statements about the hyper-K\"ahler limit of quaternionic K\"ahler geometry, and in order to progress further one must proceed example by example (see, e.g., \cite{Ambrosetti:2010tu} for a recent discussion of a specific example). Rather than taking this approach, we shall consider the rigid limit of special quaternionic K\"ahler manifolds \cite{Cecotti:1988qn}. This large class of manifolds are constructed from special K\"ahler base manifolds, and will allow us to make use of our discussion of the previous section.


\subsection{The rigid limit of special quaternionic K\"ahler manifolds}

At the tree level of
type II compactifications 
$\M_{\rm h}$ takes a special form in that its metric is entirely determined in terms of the holomorphic prepotential of a
$(2n_{\rm h}-2)$--dimensional special K\"ahler submanifold $\M_{\rm sk}$.
In this case $\M_{\rm h}$ is called ``special quaternionic K\"ahler''
and its construction is known as the c-map~\cite{Cecotti:1988qn}.

Let us denote the complex coordinates of $\M_{\rm  sk}$ by
$z^a, a=1,\ldots,\nh-1,$ its K\"ahler potential by $K^{\rm h}(z,\bar z)$ and the holomorphic prepotential by $G$. The remaining scalars in the
hypermultiplets are the dilaton $\phi$, the axion $\ax$ and  $2\nh$ real Ramond-Ramond scalars  $\xi^A, \tilde\xi_A, A=0,\ldots,\nh-1$.
An explicit form of the metric on $\M_{\rm h}$ is known as the
Ferrara-Sabharwal metric which reads \cite{Ferrara:1989ik}
\begin{equation}\begin{aligned}
  \label{qKmetric}
{\cal L}\  =\ & -(\partial \phi)^2
- e^{4\kappa\phi} (\partial \ax +\kappa\tilde\xi_A\partial\xi^A - \kappa\xi^A\partial\tilde\xi_A )^2
+g_{a\bar b} \partial z^a\partial\bar z^{\bar b}\\
&
- e^{2\kappa\phi} {\Im} {\cal M}^{AB-1} (\partial\tilde\xi + {\cal M}\partial\xi)_A{(\partial\tilde\xi + \bar{\cal M}\partial\xi)}_B\ .
\end{aligned}\end{equation}
The metric $g_{a\bar b}$ denotes the special K\"ahler metric on
$\M_{\rm  sk}$ 
which is determined in terms of a holomorphic prepotential $G(Z)$
by the relation \eqref{gdef} with $F(X)$ replaced by $G(Z)$.
The couplings ${\cal M}_{AB}$ are also determined in terms of  $G$  via
the analog of \eqref{Ndef}
\begin{equation}
  \label{Mdef}
{\cal M}_{AB} = \bar G_{AB} + 2\rmi\ \frac{{\Im}(G_{AC}) Z^C{\rm Im}(G_{BD}) Z^D}{{\Im}(G_{AC}) Z^AZ^C}\ ,
\end{equation}
where $Z^A = (\Mpl,z^a)$ are the homogeneous coordinates on $\M_{\rm sk}$.

In order to take the rigid limit we will make the same distinction as in
the vector multiplet sector, in that we split the scalars on 
$\M_{\rm  sk}$ into only
gravitationally coupled scalars with Planck-sized background values
(denoted again by $\Phi$)
and scalars in the observable sector which we continue to denote by
$z^a$ and which have small background values.
As a consequence, the rigid limit for the special K\"ahler
metric $g_{a\bar b}$ is exactly as in the previous section
and it reduces to a metric on a rigid
special K\"ahler manifold determined by a K\"ahler potential
analogous to the one given in \eqref{eq:Kr-r} 
with $\cF^{t(2)}$ defined in \eqref{Fkappa} replaced by 
$\cG^{z(2)}(\kappa\Phi_0,z,\Lambda)$
\begin{equation}\label{eq:Kr-phi}
K^{\rm h}= h(\Phi,z)+\bar h(\bar\Phi,\bar z)
-\iu\big({\delta\bar{\Phi}} \cG^{\Phi(2)}_\Phi- \delta\Phi\bar\cG^{\Phi(2)}_{\bar\Phi}\big)
- \iu \big({\bar{z}}^{\bar a} \cG^{z(2)}_a- z^a\bar\cG^{z(2)}_{\bar a}\big)\ ~,
\end{equation}
where 
${\cal G}^{\Phi(2)}_\Phi= \tfrac{i}4 g_{\Phi\bar\Phi}(\kappa\Phi_0)\,\delta\Phi^2$.
As in the observable sector of the vector multiplets we have
allowed for a non-trivial, possibly different scale $\Lambda<\Mpl$ 
also in the hypermultiplet sector.

By exactly the same reasoning as in the previous section the matrix
${\cal M}_{AB}$ defined in \eqref{Mdef} becomes block diagonal
in the rigid limit
\begin{eqnarray}
{\mathcal M}_{AB} = \left(\begin{aligned}
{\mathcal M}_{\Phi\Phi} && 0  \\
0 && \bar{{\cal G}}^{z(2)}_{ab}
\end{aligned}\right) \ ,
\end{eqnarray}
where ${\mathcal M}_{\Phi\Phi}$ is constant and includes the $A=0$ direction
while ${\bar{{\cal G}}^{z(2)}}_{ab}$ is the second derivative of 
the prepotential ${\bar{{\cal G}}^{z(2)}}$.

Finally we need to take the $\kappa\to0$ limit in the first two terms 
in \eqref{qKmetric} which merely leaves $(\partial \phi)^2+ (\partial \ax)^2$. 
Thus
the metric \eqref{qKmetric} splits into a flat part 
for the field directions  $\phi, \ax$ and the gravitationally coupled 
$\Phi$ of  the special K\"ahler manifold $\M_{\rm sk}$ together with the
corresponding RR-scalars, which we denote by $\xi^\Phi, \tilde\xi_\Phi$
and which include the $A=0$ direction.
Being flat this component is also trivially hyper-K\"ahler.
In the observable sector the hypermultiplets contain the scalars
$(z^a,\bar z^{\bar a}, \xi^a, \tilde\xi_a)$. Taking the rigid limit
of \eqref{qKmetric} in these directions one determines a K\"ahler metric
characterized  by the K\"ahler potential
\begin{equation}
{K_{\rm rc}} =  \iu(\bar z^a \cG^{z(2)}_a -z^a \bar\cG^{z(2)}_a) 
-\tfrac{1}{2} (\Im{\cal G}^{z(2)})^{-1 ab} ({C} +\bar{{C}})_a ({C} + \bar{{C}})_b \ , \label{hyperKaK}
\end{equation}
where we defined \cite{Ferrara:1989ik}
\begin{equation}\label{Cdeff}
{C}_a  =  \rmi ({\tilde \xi}_a+  {\cal G}^{z(2)}_{ab}{\xi}^b)  \ .
\end{equation}
${K_{\rm rc}} $ is  known as the K\"ahler potential of the rigid c-map
and in \cite{Cecotti:1988qn} it is shown that the corresponding metric
is hyper-K\"ahler. Flat directions can be added to
${K_{\rm rc}}$ by
replacing ${\cal G}^{z(2)}$ in \eqref{hyperKaK} by
${\cal G}^{(2)}=\tfrac{\iu}4 \tau^2+ \cG_\Phi^{\Phi (2)}+\cG^{z(2)}$, 
for $\tau=\phi+\iu\tilde\phi$, and defining $C_\Phi$ in terms of 
$\xi^\Phi,\tilde\xi_\Phi$ as in 
\eqref{Cdeff}. Note that the rigid limit that we have just described 
associates to every special quaternionic K\"ahler manifold
characterized by a prepotential $G$ 
a hyper-K\"ahler manifold characterized
by a different prepotential $\cG^{(2)}$.
This is in contrast to the recently discussed
quaternionic--hyper-K\"ahler correspondence where 
the  hyper-K\"ahler manifold is characterized by the same prepotential
$G$ (see e.g. \cite{Alexandrov:2013yva} and references therein).

We have given a prescription for the rigid limit of the special
K\"ahler and quaternionic K\"ahler geometry appearing in the
sigma-model metrics of $\cN=2$ supergravity. Using this, one could then
study the rigid limit of the scalar potential generated by gauging the
isometries of these two metrics. Indeed, it is straightforward to
check that in the rigid limit the supergravity scalar potential
reduces to that of a global $\cN=2$ supersymmetric theory. We will return to this issue, as well as the rigid limit of
spontaneously broken $N=2$ theories, in future work.  

To conclude, let us briefly mention the rigid limit of $\cN=2$ supergravity on an anti-de Sitter (AdS) background. Recently, there has been much progress in the study of rigid supersymmetry in AdS space and, in particular, it has been realised that the structures of the sigma-models are different to their flat space counterparts. For instance, in rigid $\cN=1$ AdS supersymmetry it has been shown that the usual K\"ahler manifold must have an exact  K\"aher form, and that this condition follows from the rigid limit of $\cN=1$ supergravity \cite{Adams:2011vw}. In rigid $\cN=2$ AdS supersymmetry it has been shown that one of the triplet of K\"ahler forms on the hyperK\"ahler manifold must be exact, implying that the manifold is non-compact, and that the manifold must possess an additional $SO(2)$ isometry relative to its flat space counterpart \cite{Butter:2011zt}. It would be interesting to consider how these constraints arise in the rigid limit of $\cN=2$ supergravity on an AdS background.

\subsection*{Acknowledgements}

This work was partly supported by the German Science Foundation (DFG) under the Collaborative Research Center (SFB) 676 Particles, Strings and the Early Universe. P.S. is supported by the Swiss National Science Foundation. B.E.G. is partly supported by Riset KK
 ITB, Riset Desentralisasi DIKTI-ITB, Hibah Kompetensi DIKTI, and
 DAAD. We thank Hagen Triendl for his involvement at an early stage of
 this project. We have benefited from conversations and correspondence
 with Vicente Cort\'es,  Bernard de Wit, Thomas Mohaupt, Claudio Scrucca and Antoine Van Proeyen.

\vskip 1cm


\bibliography{GLSTV}
\bibliographystyle{JHEP-2}

\providecommand{\href}[2]{#2}\begingroup\raggedright\endgroup

\end{document}